\documentclass[12pt]{article}
\usepackage[a4paper]{geometry}
\usepackage{amsfonts,amssymb,amsmath,enumerate}
\usepackage{color}
\usepackage{theorem,cite}
\usepackage{indentfirst}
\usepackage{textcomp}

\newtheorem{thm}{Theorem}[section]
\newtheorem{prop}[thm]{Proposition}
\newtheorem{lem}[thm]{Lemma}

\newtheorem{defi}[thm]{Definition}
\newcommand{\prf}{\textsl{Proof:}\ }
\newcommand{\qed}{\null \hfill {\rule{5pt}{5pt}}\\ \indent}

\theorembodyfont{\rmfamily}
\newtheorem{rmk}{Remark}[section]
\newcommand{\mb}[1]{\hspace{2.1ex}\mbox{#1}\hspace{2.1ex}}

\def\cA{{\cal A}}    \def\cB{{\cal B}}    \def\cC{{\cal C}}
        
    \def\cH{{\cal H}}    
        
\def\cM{{\cal M}}        
        
\def\cS{{\cal S}}    \def\cT{{\cal T}}

\def\fn{{\mathfrak n}}

\newcommand{\CC}{{\mathbb C}}

\begin{document}
\rightline{LAPTh-027/17}
\rightline{August 2017}
\vfill

\begin{center}

 {\LARGE  {\sffamily Back to baxterisation} }\\[1cm]

\vspace{10mm}
  
{\Large 
 N. Crampe$^{a}$\footnote{nicolas.crampe@umontpellier.fr},
 E. Ragoucy$^{b}$\footnote{eric.ragoucy@lapth.cnrs.fr}
 and M. Vanicat$^{b}$\footnote{matthieu.vanicat@lapth.cnrs.fr}}\\[.41cm] 
 $^a$ Laboratoire Charles Coulomb (L2C), UMR 5221 CNRS-Universit{\'e} de Montpellier,\\
Montpellier, F-France.
\\[.42cm]
$^{b}$ Laboratoire de Physique Th{\'e}orique LAPTh,
 CNRS and Universit{\'e} Savoie Mont Blanc.\\
   9 chemin de Bellevue, BP 110, F-74941  Annecy-le-Vieux Cedex, 
France. 
\end{center}
\vfill

\begin{abstract}
In the continuity of our previous paper \cite{Sn}, we define three new algebras, $\cA_{\fn}(a,b,c)$, $\cB_{\fn}$ and $\cC_{\fn}$, that are
close to the braid algebra. They allow to build 
solutions to the braided Yang-Baxter equation with spectral parameters. 
The construction is based on a baxterisation procedure, similar to the one used 
in the context of Hecke or BMW algebras.  

The $\cA_{\fn}(a,b,c)$ algebra depends on three arbitrary parameters, and when the parameter $a$
is set to zero, we recover the algebra $\cM_{\fn}(b,c)$ already introduced elsewhere for purpose of baxterisation. 
The Hecke algebra (and its baxterisation) can be recovered from a coset of the $\cA_{\fn}(0,0,c)$ algebra.
The algebra $\cA_{\fn}(0,b,-b^2)$ is a coset of the braid algebra.

The two other algebras $\cB_{\fn}$ and $\cC_{\fn}$ do not possess any parameter, 
and can be also viewed as a coset of the braid algebra.

\end{abstract}

\vfill
\newpage
\pagestyle{plain}
\section*{Introduction} 

The Yang-Baxter equation with spectral parameters and its solutions, the R-matrices, are crucial to define and study the quantum integrable systems.
Indeed, the R-matrices allow one to construct the generating function of the conserved commuting quantities of the integrable model.
Thus, the computation of the explicit R-matrices is the first step to construct new integrable models.
Unfortunately, the Yang-Baxter equation corresponds to functional coupled equations difficult to solve.

To overcome this problem, different techniques have been proposed. 
In particular, V.F.R. Jones \cite{jones}, in a framework of knot theory, suggests to obtain solutions of the YBE with spectral parameter 
from representations of some algebras as e.g. some quotients of the braid group. This method is called Baxterisation.
The importance of this method relies on the fact it is possible to baxterise the Hecke algebra, the Temperley--Lieb algebra or the Birman--Murakami--Wenzl algebra 
\cite{Jim86,Isaev}. Then, these results have been generalized to obtain other solutions of the Yang-Baxter equation see e.g. 
\cite{CGX,ZGB,YQL,BM2000,ACDM,KMN,FRR13,Sn,theseMatthieu}.

In this short note, we propose a generalisation of this procedure based on new algebras.
In section \ref{sec:algebra}, we define and study these different new algebras which are used in section \ref{sec:baxt} to
provide baxterisations. The main results are encompassed in the theorem \ref{th:baxtA}.

\section{Algebras \label{sec:algebra}}

In this section, we introduce different algebras and study some of their properties. 
All these algebras are used to construct R-matrices,  through a process called baxterisation (see section \ref{sec:baxt}). 
Before introducing them, we remind the definition of two well-known algebras:
\begin{defi}
The Braid algebra  is generated by the generators $\sigma_i$, $i=1,2,...,\fn-1$, 
and submitted to the following relations:
\begin{itemize}
 \item the locality relation, for $i,j=1,2,...,\fn-1$ and $|i-j|>1$,
 \begin{eqnarray}
&& [\sigma_i,\sigma_j]=0  \label{eq:com}
\end{eqnarray}  
\item the braid relation, for $i=1,2,...,\fn-2$, 
\begin{eqnarray}
&& \sigma_i\sigma_{i+1}\,\sigma_i=\sigma_{i+1}\,\sigma_i\,\sigma_{i+1}. \label{eq:braid}
\end{eqnarray}
\end{itemize}
\end{defi}
\begin{defi}
The Hecke algebra $\cH_{\fn}(q)$, $q\in\CC$,  is generated by the generators $\sigma_i$, $i=1,2,...,\fn-1$, 
and submitted to the relations \eqref{eq:com}, \eqref{eq:braid} and
 \begin{eqnarray}
&& \label{eq:Hecke}
(\sigma_i-1)(\sigma_i-q)=0\,,\quad i=1,2...,\fn-1.
\end{eqnarray}  
\end{defi}
Interesting enough, although the braid algebra is well-known for its relation to knot theory, 
no direct baxterisation can be defined for the braid algebra. 
However, such a procedure is well-known for the Hecke algebra, that is a coset of the braid algebra. 
More generally, a classification of all baxterisations that can be built on cosets of the braid algebra 
is far from been known, neither the class of algebras that allow to built some baxterisation. 
We view the algebras we present hereafter as some building blocks toward such classifications.

\subsection{The algebra $\cA_{\fn}(a,b,c)$}

\begin{defi}
The  algebra $\cA_{\fn}(a,b,c)$ is generated by the generators $\sigma_i$, $i=1,2,...,\fn-1$, 
and submitted to the locality relation \eqref{eq:com} and the following relations
 \begin{eqnarray}
&& [\sigma_i^2,\sigma_{i+1}] = [\sigma_i,\sigma_{i+1}^2] \label{eq:aa1}\\
&&{[\sigma_{i+1}\sigma_i,\sigma_{i}+\sigma_{i+1}]}  = a(\sigma_{i}^3-\sigma_{i+1}^3)
+b(\sigma_{i}^2-\sigma_{i+1}^2)-c(\sigma_{i}-\sigma_{i+1}) \label{eq:aa2}  \\
&&a(\sigma_i\sigma_{i+1}^3-\sigma_i^3\sigma_{i+1})=-b( \sigma_{i+1}^2\sigma_i-\sigma_{i+1}\sigma_i^2 )  \label{eq:aa5}\\
&& a(\sigma_{i+1}^3\sigma_i-\sigma_{i+1}\sigma_i^3)=-b( \sigma_{i+1}^2\sigma_i-\sigma_{i+1}\sigma_i^2 )  \label{eq:aa3} \\
&& a^2(\sigma_{i+1}^4\sigma_i-\sigma_{i+1}\sigma_i^4)=(b^2+ac)( \sigma_{i+1}^2\sigma_i-\sigma_{i+1}\sigma_i^2 )   \label{eq:aa4} 
\end{eqnarray}  
\end{defi}

\begin{prop}
 If $a\neq 2$, relation \eqref{eq:aa5} is implied by relations \eqref{eq:aa1}, \eqref{eq:aa2} and \eqref{eq:aa3}.
\end{prop}
\prf
Taking the following combination of relations \eqref{eq:aa1}, \eqref{eq:aa2} and \eqref{eq:aa3}
\begin{equation}
(a-2)\eqref{eq:aa3}+a [\sigma_i- \sigma_{i+1} , \eqref{eq:aa2}]
-a(2\sigma_i+\sigma_{i+1})\eqref{eq:aa1}-a\eqref{eq:aa1}(\sigma_i+2\sigma_{i+1})-ab\eqref{eq:aa1}\;,\qquad
\end{equation}
we find relation \eqref{eq:aa5} multiplied by $a-2$.
\qed

\begin{prop}
We have the following correspondences between algebras.
\begin{itemize}
\item The Hecke algebra $\cH_{\fn}(q)$ is the coset of the $\cA_{\fn}(0,0,-q)$ algebra by the relation \eqref{eq:Hecke}.
\item The algebra $\cA_{\fn}(0,b,c)$  for $b\neq 0$ is isomorphic to the algebra $\cM_{\fn}(b,c)$ introduced in \cite{theseMatthieu} as a generalisation of the algebra appearing in \cite{H33}.
\item The algebra $\cA_{\fn}(0,b,-b^2)$ with $b\neq0$ is the coset of the braid algebra by the relations 
\begin{eqnarray*}
&&\tau_{i+1}^2 \tau_{i} -\tau_{i+1} \tau_{i}^2 =b^2(\tau_{i} -\tau_{i+1})-b(\tau_{i}^2 -\tau_{i+1}^2),
\\ 
&&\tau_{i}^2 \tau_{i+1} -\tau_{i} \tau_{i+1}^2 =b^2(\tau_{i+1} -\tau_{i})-b(\tau_{i+1}^2 -\tau_{i}^2).
\end{eqnarray*}
\end{itemize}
\end{prop}
\prf Direct calculation. In the last construction, the generators $\tau_i$ of the braid algebra are related to the generators $\sigma_i$ 
of the $\cA_{\fn}$ algebra through $\sigma_i=\tau_i-b$.
\qed

\begin{prop}
The mapping $\sigma_j\to\sigma_{\fn-j}$, $j=1,2,...,\fn-1$, defines an automorphism of the algebra $\cA_{\fn}(a,b,c)$.
\end{prop}
\prf Relation \eqref{eq:aa1} implies that the set of relations \eqref{eq:aa2}-\eqref{eq:aa3} is invariant under the replacement. 
These relations also imply that 
$ [\sigma_i^p,\sigma_{i+1}] = [\sigma_i,\sigma_{i+1}^p] $, $\forall p$, which makes the last defining relation invariant too. \qed

\subsection{The algebras $\cB_{\fn}$ and $\cC_{\fn}$}

\begin{defi}
The  algebra $\cB_{\fn}$ is generated by the generators $\sigma_i$, $i=1,2,...,\fn-1$, 
and submitted to the locality relation \eqref{eq:com}, the braid relation \eqref{eq:braid} and the following relations
 \begin{eqnarray}
&& \sigma_i^2\sigma_{i+1} - \sigma_i\sigma_{i+1}^2 = \sigma_{i}^2 -\sigma_{i+1}^2 +\sigma_{i+1} -\sigma_i \label{eq:bb2}\\
&& \sigma_{i+1}^3\sigma_i-\sigma_{i+1}\sigma_i^3= \sigma_{i+1}^2\sigma_i-\sigma_{i+1}\sigma_i^2
  +\sigma_{i+1}^3-\sigma_i^3-\sigma_{i+1}^2+\sigma_i^2
  \qquad\label{eq:bb3} \\
&& \sigma_{i+1}^4\sigma_i-\sigma_{i+1}\sigma_i^4 = \sigma_{i+1}^2\sigma_i-\sigma_{i+1}\sigma_i^2  
  +\sigma_{i+1}^4-\sigma_i^4-\sigma_{i+1}^2+\sigma_i^2
\quad\label{eq:bb4} 
\end{eqnarray}  
\end{defi}
We note that this algebra is a quotient of the Braid algebra by the relations \eqref{eq:bb2}-\eqref{eq:bb4}.
\begin{rmk}
The coset of the $\cB_{\fn}$ algebra by relations \eqref{eq:Hecke} is the Hecke algebra $\cH_{\fn}(q)$.
\end{rmk}

\begin{rmk}
The coset of the $\cB_{\fn}$ algebra by the relations
$\sigma_{i+1}\sigma_i^2 - \sigma_{i+1}^2\sigma_i = \sigma_{i}^2 -\sigma_{i+1}^2 +\sigma_{i+1} -\sigma_i$, $i=1,2,...,\fn-1$
is isomorphic to $\cA_{\fn}(0,b,-b^2)$ (for $b\neq 0$) via the homomorphism $\cB_{\fn}\rightarrow \cA_{\fn}(0,b,-b^2)$, $\sigma_i\mapsto \frac{1}{b}\sigma_i+1$.
\end{rmk}

For purpose of baxterisation, we introduce another algebra:
\begin{defi}
The  algebra $\cC_{\fn}$ is generated by the generators $\tau_i$, $i=1,2,...,\fn-1$, 
and submitted to the locality relation \eqref{eq:com}, the braid relation \eqref{eq:braid} and the following relations
 \begin{eqnarray}
&& \tau_{i+1}^2\tau_{i} - \tau_{i+1}\tau_{i}^2 = \tau_{i+1}^2 -\tau_{i}^2 +\tau_{i} -\tau_{i+1} \label{eq:cc2}\\
&& \tau_{i}^3\tau_{i+1}-\tau_{i}\tau_{i+1}^3= \tau_{i}^2\tau_{i+1}-\tau_{i}\tau_{i+1}^2
  +\tau_{i}^3-\tau_{i+1}^3-\tau_{i}^2+\tau_{i+1}^2
  \qquad\label{eq:cc3} \\
&& \tau_{i}^4\tau_{i+1}-\tau_{i}\tau_{i+1}^4 = \tau_{i}^2\tau_{i+1}-\tau_{i}\tau_{i+1}^2  
  +\tau_{i}^4-\tau_{i+1}^4-\tau_{i}^2+\tau_{i+1}^2
\quad\label{eq:cc4} 
\end{eqnarray}  
\end{defi}
This algebra is obviously isomorphic to the $\cB_{\fn}$ algebra through the correspondence $\tau_j=\sigma_{\fn-j}$. 

\subsection{Representations}
We classify the scalar representations of the algebras $\cA_{\fn}(a,b,c)$, $\cB_{\fn}$ and $\cC_{\fn}$ and give some irreducible 2-dimensional representations.

\begin{prop}\label{scalar}
The algebras $\cA_{\fn}(a,b,c)$, $\cB_{\fn}$ and $\cC_{\fn}$ admit two classes of scalar representations:
\begin{eqnarray*}
&& (i)\ \sigma_i=\lambda\,,\ \forall i \mb{where}\lambda\in\CC.\\ 
&&(ii)\ \sigma_{i}\in\{0,\lambda_0^-,\lambda_0^+\}\mb{with}
\begin{cases}\lambda_0^+=\lambda_0^-=1\mb{for}\cA_{\fn}(0,b,c)\,,\ \cB_{\fn} \mb{and} \cC_{\fn}\\[1ex]
\lambda_0^\pm=\frac1{2a}\Big({-b\pm\sqrt{b^2+4ac}}\Big)\mb{for}\cA_{\fn}(a,b,c)\,,\  a\neq0.\end{cases} 
\end{eqnarray*}
In the case $(ii)$, the choice for $\sigma_i$  in the set $\{0,\lambda_0^-,\lambda_0^+\}$ is made independently for the different values of $i=1,...,\fn-1$.
\end{prop}
\prf We set $\sigma_i=\lambda_i\in\CC$ in the different algebras. We start with the algebras $\cA_{\fn}(0,b,c)$, $\cB_{\fn}$ and $\cC_{\fn}$.
The relations \eqref{eq:aa1} or \eqref{eq:braid} imply that we have $\lambda_i\lambda_{i+1}(\lambda_i-\lambda_{i+1})=0$. 
This implies that either $\lambda_i=\lambda_{i+1}=\lambda$, which corresponds to $(i)$, or $\lambda_i\lambda_{i+1}=0$, which corresponds to the possibility $(ii)$. The examination of the other algebraic relations shows that $(i)$ is valid for all values of $\lambda$, while $(ii)$ is consistent only if the non-vanishing $\lambda_i$ are equal to 1.

As far as $\cA_{\fn}(a,b,c)$ is concerned, the defining relations in the case of a scalar representation reduce to
\begin{eqnarray}
&&\lambda_i(a\lambda_i^2+b\lambda_i-c)=\lambda_{i+1}(a\lambda_{i+1}^2+b\lambda_{i+1}-c)
\\
&&\lambda_i\lambda_{i+1}(\lambda_i-\lambda_{i+1})\big(a(\lambda_i+\lambda_{i+1})+b\big)=0\label{eq:truc1}
\\
&&\lambda_i\lambda_{i+1}(\lambda_i-\lambda_{i+1})\big(a^2(\lambda_i^2+\lambda_{i+1}\lambda_i+\lambda_{i+1}^2)-b^2-ac\big)=0.\label{eq:truc2}
\end{eqnarray}
 When $a=0$, we are back to the previous case, so that we can suppose that $a\neq0$.
The resolution of these equations leads to the cases presented above. In particular, the values $\lambda_0^\pm$ are the roots of the equation $a\lambda^2+b\lambda-c$, and the couple $(\lambda_i,\lambda_{i+1})=(\lambda_0^\pm,\lambda_0^\mp)$ cancels the last factor in \eqref{eq:truc1} and \eqref{eq:truc2}. 
\qed

In the next section we will be mainly interested in the algebras $\cA_{3}(a,b,c)$, $\cB_{3}$ and $\cC_{3}$ 
in the context of the Yang-Baxter equation. The following
lemma ensures that they have non trivial matricial representations, opening the door for the construction of physically relevant integrable Hamiltonians. 
Since $\cB_{3}$ and $\cC_{3}$ are isomorphic, we focus on 
the algebras $\cA_{3}$ and $\cB_{3}$.
\begin{lem}The algebra $\cA_3(a,b,c)$ admit non-trivial two-dimensional representations. An example is given by
$$
\sigma_1=\left(\begin{array}{cc} 0 & c \\ 0 & 0\end{array}\right)
\mbox{ ; } 
\sigma_2=\left(\begin{array}{cc} \mu &  
-\mu^2 
\\ 1 & -\mu\end{array}\right)
$$
where $\mu\in\CC$ is a free parameter.

The algebra $\cB_3$ admits non-trivial two-dimensional representations. An example is given by
$$
\sigma_1=\left(\begin{array}{cc} \nu\mu & 0 \\ \nu & 1\end{array}\right)
\mb{ and } 
\sigma_2=\left(\begin{array}{cc} 1 &  -\nu \\0 & \nu\mu\end{array}\right)\,,\quad \nu,\mu\in\CC.
$$
\end{lem}

\prf
One shows by direct calculation that the matrices indeed obey the defining relations of the algebras.\qed
%

\section{Baxterisation \label{sec:baxt}}

We  show in this section that the algebras we have introduced allow to obtain solutions of the Yang-Baxter equation.
\subsection{Presentation of the main result}
The method to get solutions of the Yang-Baxter equation is called baxterisation \cite{jones}.
Indeed, let us define the following algebraic elements 
\begin{eqnarray}
 \check R_i(x,y)=(1-f(x,y) \sigma_i)(1-f(y,x) \sigma_i)^{-1}\;,\quad i=1,2,\label{eq:Rmatrix}
 \end{eqnarray}
where the inverse $(1-f(y,x) \sigma_i)^{-1}$ is understood as a formal series expansion $\sum_{p\geq0} (f(y,x)\sigma_i)^p$.
 For a given rational function $f(x,y)$, the elements $\sigma_1$ and $\sigma_2$ have to obey suitable algebraic relations in order for 
$ \check R_i(x,y)$ to satisfy
 the braided Yang-Baxter equation
\begin{equation}
   \check R_{1}(x,y) \check R_{2}(x,z)\check R_{1}(y,z)=\check R_{2}(y,z)\check R_{1}(x,z)\check R_{2}(x,y)\;.\label{eq:YBE}
   \end{equation}
When $f(x,y)=x/y$, and $\sigma_i$, $i=1,2$ belong to the Hecke or BMW algebras,
the form \eqref{eq:Rmatrix}  just corresponds to the baxterisation
for these algebras  (see e.g. \cite{IO}).

The form \eqref{eq:Rmatrix} appears also naturally in the context of the coordinate Bethe ansatz \cite{H33} with $f(x,y)=\frac{y}{mxy+n}$.
The cases $f(x,y)=x$ or $f(x,y)=y$ have been treated in \cite{Sn}: they rely on two isomorphic algebras $\cS_\fn$ and $\cT_\fn$ described in \cite{Sn}.
The case $f(x,y)=\frac{x+\lambda xy}{1-\mu xy}$ has been studied in \cite{theseMatthieu} and makes appear the algebra $\cM_\fn(b,c)\equiv\cA_\fn(0,b,c)$.
   
In the following  theorem, we give the explicit forms of $f(x,y)$  that correspond to the algebras $\cA_{\fn}$,  
$\cB_{\fn}$ and $\cC_{\fn}$. 
\begin{thm}\label{th:baxtA}
Let $\check R(x,y)$ defined by \eqref{eq:Rmatrix} where the function $f(x,y)$ takes one of the three following forms
\begin{eqnarray}
&&\textit{(i)}\quad f(x,y)=\frac{\alpha_1 x+\alpha_2 y+b xy}{1+c xy}\;,\mbox{ with }\ |\alpha_1- \alpha_2|=1 \mbox{ and } \alpha_1 \alpha_2=a,
\label{eq:fA}\\
&&\textit{(ii)}\quad  f(x,y)=\frac{(1+y)x}{1+x}\;,
\\
&&\textit{(iii)}\quad f(x,y)=\frac{(1+x)y}{1+y}\;.
\end{eqnarray}
Then $\check R(x,y)$
satisfies the braided Yang--Baxter equation \eqref{eq:YBE} when
$\sigma_1$ and $\sigma_2$ belong to the algebra $\cA_{3}(a,b,c)$ (case \textit{(i)}), $\cB_{3}$ (case \textit{(ii)}), or $\cC_{3}$ (case \textit{(iii)}).
\end{thm}

\subsection{Proof of theorem \ref{th:baxtA}}
Before proving the theorem, we introduce two lemmas that we will need. 
\begin{lem}\label{lem:pra}
If 
$\sigma_1$ and $\sigma_2$ satisfy $\cA_{3}(a,b,c)$ for $a\neq 0$, then the following relations hold 
\begin{eqnarray}
 &&a\  (\sigma_2^3 \sigma_1^2-\sigma_2^2 \sigma_1^3) =-c\ (\sigma_2^2\sigma_1-\sigma_2\sigma_1^2)=a\  (\sigma_1^2\sigma_2^3 -\sigma_1^3\sigma_2^2 ) \label{eq:rel1}\\
 &&\sigma_2 H_1(z)-H_2(z)\sigma_1=zh(z)(\sigma_2^2\sigma_1-\sigma_2\sigma_1^2)=H_1(z)\sigma_2-\sigma_1 H_2(z) 
 \label{eq:rel2}\\
 &&H_2(v)H_1(z)-H_2(z)H_1(v)=(v-z)h(z)h(v)(\sigma_2^2\sigma_1-\sigma_2\sigma_1^2)  \label{eq:rel4}\\
 &&\sigma_1H_2(v)H_1(z)-H_2(z)H_1(v)\sigma_2=\frac{a}{zv}(\sigma_2-\sigma_1)+\frac{c zv-bv-a}{a}h(z)h(v)
 (\sigma_2^2\sigma_1-\sigma_2\sigma_1^2)\nonumber\\
 &&\qquad\qquad+\frac{a}{v(v-z)h(v)}(H_1(v)-H_2(v))-\frac{a}{z(v-z)h(z)}(H_1(z)-H_2(z)) \label{eq:rel5}
\end{eqnarray}
where
\begin{eqnarray}
 H_i(z)=\sum_{\ell=0}^{\infty}\sigma_i^{\ell+1} z^\ell=\sigma_i(1-z\sigma_i)^{-1}
 \label{eq:defH}\mb{and}
 h(z)&=&\frac{a}{c z^2-b z -a}\;.
\end{eqnarray}
\end{lem}
\prf 
Considering the expression $\sigma_2\eqref{eq:aa3}+\eqref{eq:aa3}\sigma_1$, we get
\begin{equation}
 a(\sigma_2^3\sigma_1^2 - \sigma_2^2\sigma_1^3)+a(\sigma_2^4\sigma_1 -\sigma_2\sigma_1^4 )=-b(\sigma_2^3\sigma_1 - \sigma_2\sigma_1^3)\;.
\end{equation}
Then, using \eqref{eq:aa3} and \eqref{eq:aa4}, we prove the first equality of \eqref{eq:rel1} (we have used here that $a\neq 0$). 
The second one is proved similarly.

After multiplication on the left by $(1-z\sigma_2)$ and on the right by$(1-z\sigma_1)$, the first equality of \eqref{eq:rel2} is equivalent to 
\begin{equation}
-\sigma_2^2 \sigma_1+\sigma_2 \sigma_1^2 =h(z)\Big(\sigma_2^2 \sigma_1-\sigma_2 \sigma_1^2-z(\sigma_2^3 \sigma_1-\sigma_2 \sigma_1^2)
+z^2(\sigma_2^3 \sigma_1^2-\sigma_2^2 \sigma_1^3)\Big)\;.
\end{equation}
This last relation is proved directly using relations \eqref{eq:rel1} and \eqref{eq:aa3} and the explicit form \eqref{eq:defH} of the function $h$.
The second equality of \eqref{eq:rel2} is proved in the same way.

Relation \eqref{eq:rel4} is equivalent to (multiplying on the left by $(1-v\sigma_2)$ and on the right by $(1-v\sigma_1)$)
\begin{equation}\label{eq:pr2}
\sigma_2 H_1(z)(1-v\sigma_1)-(1-v\sigma_2)H_2(z)\sigma_1=(v-z)h(z)h(v) (1-v\sigma_2)(\sigma_2^2\sigma_1-\sigma_2\sigma_1^2)(1-v\sigma_1) \;.
\end{equation}
The R.H.S. is simplified by using relations \eqref{eq:rel1} and \eqref{eq:aa3} and the L.H.S. is simplified by remarking 
that $\sigma_iH_i(z)=\frac{1}{z}(H_i(z)-\sigma_i)$ and by using \eqref{eq:rel2}. Then, with the explicit form of the function $h$, 
it is easy to show that \eqref{eq:pr2} holds which proves \eqref{eq:rel4}.

To prove relation \eqref{eq:rel5}, we show that the following relation holds
\begin{eqnarray}
 &&v(v-z)h(z)h(v)(1-v\sigma_1)(1-z\sigma_2)(\sigma_2^2\sigma_1-\sigma_2\sigma_1^2)(1-z\sigma_1)(1-v\sigma_2)\\
 &=&(1-v\sigma_1)(1-z\sigma_2)\left(\sigma_1-\frac{1}{z}-\frac{av}{z(z-v)h(z)}+\frac{v(1+a+bz)}{z}\sigma_2+av\sigma_2^2\right)\nonumber\\
 &-&\left(\sigma_2-\frac{1}{z}-\frac{av}{z(z-v)h(z)}+\frac{v(1+a+bz)}{z}\sigma_1+av\sigma_1^2\right)(1-z\sigma_1)(1-v\sigma_2),\nonumber
\end{eqnarray}
by expanding all the factors and by using the defining relations of $\cA_\fn$. Then, we multiply it on the left by $(1-v\sigma_1)^{-1}(1-z\sigma_2)^{-1}$
and on the right by $(1-z\sigma_1)^{-1}(1-v\sigma_2)^{-1}$ and we use that $(1-x\sigma_i)^{-1}=xH_i(x)+1$, $\sigma_i(1-x\sigma_i)^{-1}=H_i(x)$ 
and $\sigma_i^2(1-x\sigma_i)^{-1}=\frac{1}{x}(H_i(x)-\sigma_i)$ to obtain  \eqref{eq:rel5} after rearrangement of all the terms.
\qed
\begin{lem}\label{lem:prb}
If 
$\sigma_1$ and $\sigma_2$ satisfy $\cB_{3}$, then the following relations hold 
\begin{eqnarray}
 && \sigma_2^3 \sigma_1^2-\sigma_2^2 \sigma_1^3 =\sigma_2^3-\sigma_1^3+\sigma_1^2-\sigma_2^2 \label{eq:relb1}\\
 &&\sigma_2 H_1(z)-H_2(z)\sigma_1=\frac{z}{z-1}(\sigma_2^2\sigma_1-\sigma_2\sigma_1^2+\sigma_1^2-\sigma_2^2)+\sigma_2-\sigma_1+H_1(z)-H_2(z)\qquad
 \label{eq:rel2b}\\
 &&\sigma_1 H_2(z)-H_1(z)\sigma_2=\frac{1}{z-1}(\sigma_2-\sigma_1)-H_1(z)+H_2(z)\qquad
 \label{eq:rel2bb}\\
 &&H_2(v)H_1(z)-H_2(z)H_1(v)=\frac{v-z}{(v-1)(z-1)}(\sigma_2^2\sigma_1-\sigma_2\sigma_1^2+\sigma_1^2-\sigma_2^2-\sigma_1+\sigma_2) \nonumber \\
 &&\qquad\qquad+\frac{1}{v-1}(H_2(z)-H_1(z))-\frac{1}{z-1}(H_2(v)-H_1(v))\label{eq:rel4b}\\
 &&\sigma_1H_2(v)H_1(z)-H_2(z)H_1(v)\sigma_2=\frac{vz-z+1}{z(z-1)(v-1)}(\sigma_2-\sigma_1)\nonumber\\
 &&\qquad\qquad+\frac{v}{(v-1)(z-1)} (\sigma_2^2\sigma_1-\sigma_2\sigma_1^2+\sigma_1^2-\sigma_2^2)\nonumber\\
 &&\qquad\qquad+\frac{v-z+1}{(v-z)(z-1)}(H_1(v)-H_2(v))-\frac{v}{z(v-1)(v-z)}(H_1(z)-H_2(z)) \label{eq:rel5b}
\end{eqnarray}
where $H_i=\sigma_i(1-z\sigma_i)^{-1}=\sum_{\ell=0}^{\infty}\sigma_i^{\ell+1} z^\ell$.
\end{lem}
\prf The proof of this lemma is similar to the one of the lemma \ref{lem:pra}.
\qed
\medskip

We are now in position to prove the theorem \ref{th:baxtA}.

The subcase $(i)$ with $a=0$ was proven in \cite{theseMatthieu}, so that we prove $(i)$ supposing that $a\neq0$.
By multiplying the Yang-Baxter equation \eqref{eq:YBE} on the left by $(1-f(x_2,x_1) \sigma_1)$ and on the right by $(1-f(x_2,x_1) \sigma_2)$, 
it can be written equivalently as follows
\begin{equation}
(1-f_{12} \sigma_1) M(x_1,x_2) (1-f_{21} \sigma_2)=
(1-f_{21} \sigma_1) M(x_2,x_1) (1-f_{12} \sigma_2)\;,
\end{equation}
where we have used the notation $f_{ij}=f(x_i,x_j)$ and
\begin{equation}
 M(x_1,x_2)=\check R_2(x_1,x_3) \check R_1(x_2,x_3)\;.
\end{equation}
Then, the Yang-Baxter equation can be written
\begin{eqnarray}
 &&M(x_1,x_2)-M(x_2,x_1)+f_{12}f_{21}\sigma_1( M(x_1,x_2)-M(x_2,x_1) )\sigma_2\nonumber\\
 &=&f_{12}(\sigma_1 M(x_1,x_2)-M(x_2,x_1)\sigma_2)+f_{21}( M(x_1,x_2)\sigma_2-\sigma_1M(x_2,x_1) )\;.
 \label{eq:ybeb}
\end{eqnarray}
By remarking that
\begin{equation}
 \check R_i(x_1,x_2)=1-(f_{12}-f_{21})H_i( f_{21} )\quad\text{(where } H_i \text{ is defined in \eqref{eq:defH})}\;,
\end{equation}
one gets
\begin{eqnarray}
&& M(x_1,x_2)-M(x_2,x_1)=(f_{31}-f_{13})(f_{32}-f_{23})\Big(H_2(f_{31})H_1(f_{32})-H_2(f_{32})H_1(f_{31})\Big)\nonumber\\
 &&\hspace{2cm}+(f_{13}-f_{31})(H_1(f_{31})-H_2(f_{31}))+(f_{32}-f_{23})(H_1(f_{32})-H_2(f_{32}))\;.
\end{eqnarray}
Then, thanks to \eqref{eq:rel4}, one gets
\begin{eqnarray}
&& M(x_1,x_2)-M(x_2,x_1)=(f_{31}-f_{13})(f_{32}-f_{23})(f_{31}-f_{32})h(f_{31})h(f_{32}) (\sigma_2^2\sigma_1-\sigma_2\sigma_1^2)\nonumber\\
&&\hspace{2cm}+(f_{13}-f_{31})(H_1(f_{31})-H_2(f_{31}))+(f_{32}-f_{23})(H_1(f_{32})-H_2(f_{32}))\;.
\end{eqnarray}
Similarly, using the relations of the lemma \ref{lem:pra} and the previous equation, we express
$\sigma_1( M(x_1,x_2)-M(x_2,x_1) )\sigma_2$, $\sigma_1 M(x_1,x_2)-M(x_2,x_1)\sigma_2$ and $ M(x_1,x_2)\sigma_2-\sigma_1M(x_2,x_1)$
in terms of $\sigma_2^2\sigma_1-\sigma_2\sigma_1^2$, $H_1(f_{31})-H_2(f_{31})$, $H_1(f_{32})-H_2(f_{32})$, $\sigma_1-\sigma_2$.
Therefore, we can express relation \eqref{eq:ybeb} in terms of 
these 4 algebraic elements.
Finally, we prove that each coefficient in front of these algebraic elements vanishes, which concludes the proof of \textit{(i)}.

The part \textit{(ii)} is proven similarly to the part \textit{(i)} using the lemma \ref{lem:prb}. 
As already mentioned, the algebra $\cC_3$ is isomorphic to $\cB_3$, 
so that the proof of \textit{(iii)} is similar to the one of \textit{(ii)}.
\qed

\section{Conclusion}
In this note, we have proposed three new algebras that allow a baxterisation procedure, to get (possibly new) solutions to the braided Yang-Baxter equation with spectral parameter. 
From this perspective, these algebras are on equal footing with the Hecke, the BMW, or the cyclotomic algebras used in physics to define integrable models. They  also connect to the $\cS_\fn$ and $\cT_\fn$ algebras, introduced recently in \cite{Sn}. 
Clearly, the study of the representation theory for these new algebras is a point to be developed if one wants to elaborate from them new integrable models relevant to physics problems. 

More generally, a complete classification of such algebras allowing baxterisation would be a desirable (although technically complicated) task to acquire. One can also observe that some of the algebras are quotient of the braid algebra. A classification of the cosets of the braid algebra allowing the baxterisation procedure would be also a very interesting point to study. Finally, one can also wonder whether there exist an algebra that would encompass the remaining   algebras,
and describing them as quotients of this ``bigger" algebra.

\end{document}